\documentstyle[pra,aps,floats,twocolumn]{revtex}

\newcommand{\cH}{{\cal H}}
\newcommand{\1}{{\bf 1}}

\begin{document}
\draft
\title{Fidelity and concurrence of conjugated states}
\author{Armin Uhlmann}
\address{Institut f\"ur Theoretische Physik, Universit\"at Leipzig}
\maketitle
\begin{abstract}
We prove some new properties of fidelity (transition probability)
and concurrence, the latter defined by a straightforward extension
of Wootters' notation. Choose a conjugation and consider the
dependence of fidelity or of concurrence on conjugated pairs of
density operator. These functions turn out to be concave or
convex roofs. Optimal decompositions are constructed. Some
applications to two- and tripartite systems illustrate
the general theorems.
\end{abstract}

%

\section{introduction}

In Physics antilinearity is well known from symmetries
with time reversal operations \cite{Wigner},
from second quantization, and
from representation theory of groups and algebras.
Quantum information theory offers several new
interesting applications of antilinearity. In the present
paper we are concerned with one of them.
Antilinear operators are intrinsically non--local: One cannot
tensor them consistently with the identity operator.
They do not share the privilege of linear operators \cite{Jo98}
to allow execution in one part of a bipartite system
while ''doing nothing'' in the other one. It seems, therefore,
quite natural to use antilinear operators to describe or to
estimate effects of entanglement. Indeed, Hill and Wootters in
\cite{HW97} and Wootters in \cite{Woo97} used a particular
conjugation, the {\em Hill--Wootters conjugation}, in order
to get an explicit expression for the {\em entanglement of formation}
for two qubits. Their papers are the very starting point for
the present contribution. I tried to distil a
general method out of their proofs, and to construct explicitly
the relevant optimal decompositions.
The entanglement of formation concept is due to
Bennett et all \cite{BDSW96}. Also a peculiar basis, the
{\em magic basis}, with which one can define the
Hill-Wootters conjugation, is already in that important paper.

In the 2-qubit case the entanglement of formation is a function
of just one other quantity, called {\em (pre)concurrence},
 \cite{Woo97}, and the same optimal decomposition of
a state into pure ones can be used to calculate its entanglement
of formation and its concurrence. In this form the statement becomes
wrong for general states of a bipartite system different from the
$2 \times 2$ case. But for density operators of rank two
similar results seem not out of range.\\
However, concurrence seems to be an interesting quantity in
its own: It can be defined in higher dimensional
Hilbert spaces and with respect to any conjugation $\Theta$
by an explicit expression (section II) which will be called
$\Theta$-concurrence. It is a convex function on the state
space (section III), and it is a roof (see section V). Optimal
decompositions can be obtained (section IV) in a constructive
manner, adding some news even for the 2-qubit case. Generally,
the length of an optimal decompositions will be
the smallest power of two which  exceeds the dimension of the
Hilbert space.\\
The idea, pointing to the definition of
$\Theta$-concurrences,
can be extended to another interesting quantity, to the
{\em fidelity},  the square root of the
{\em transition probability} \cite{fidel}. $\Theta$-fidelity
as defined in section II, turns out to be a concave
roof. Optimal decompositions can be gained similarly.\\
The main proofs are in sections III and IV. Section V
is devoted to the roof concept, \cite{Uh98a}, an interesting
tool if combined with convexity or concavity.
The last section contains some applications, mainly of
$\Theta$-concurrences. There are conjugations in multipartite
systems such that a non-zero $\Theta$-concurrence indicates
inseparability. It is illustrated for bipartite (example 1)
and for the 3-qubit systems (example 3).
In a $2 \times n$ bipartite system there is the possibility
to bound entanglement of formation from below by the aid of
$\Theta$-concurrences (example 2). After extending the
method slightly (theorem 5) to a larger class of antilinear
operators, example 4 treats $\Theta$-fidelity and -concurrence
on some 2-dimensional subspaces of the 2-qubit system. Though
the result is essentially known for the concurrence, \cite{HW97},
it explains a part of the method.\\

Now I shortly call attention to some notations and rules,
connected with antilinearity, to prepare what follows below.
An {\em antilinear} operator $\vartheta$
acting on an Hilbert space ${\cH}$ satisfies by definition
$$
\vartheta \, ( a_1 \psi_1 + a_2 \psi_2 ) = a_1^* \psi_1
+ a_2^* \psi_2
$$
If $\psi$ is an eigenvector of $\vartheta$ with eigenvalue
$\lambda$, $\epsilon \psi$ is an eigenvector with eigenvalue
$\epsilon^{-2} \lambda$ for all unimodular numbers $\epsilon$.
The fact that the eigenvalues of an antilinear operator fill
some circles in the complex plain will be used in the estimations
of section III.
The product of two antilinear operators becomes
linear, the product of an of antilinear operator and a linear
one remains antilinear.
The {\em adjoint} (or Hermitian adjoint), $\vartheta^{\dagger}$,
of  an antilinear operator  $\vartheta$
is determined by the relation
$$
\langle \psi, \vartheta^{\dagger} \varphi \rangle =
\langle \varphi, \vartheta \psi \rangle
$$
for all $\psi, \, \varphi \in \cH$.
Notice $(\vartheta^{\dagger})^{\dagger} = \vartheta$.
The standard rule $(AB)^{\dagger} = B^{\dagger}A^{\dagger}$
for linear operators remains valid if one or both operators
are replaced by antilinear ones. In particular,
with a complex number $a$ and antilinear $\vartheta$ one gets
$(a \vartheta)^{\dagger} =
\vartheta^{\dagger} a^* = a \vartheta^{\dagger}$, i.~e. taking the
adjoint is a linear procedure for antilinear operators.
It follows:
The set of operators which are antilinearly Hermitian
(antilinearly self-adjoint),
$\vartheta = \vartheta^{\dagger}$, is a linear space of
dimension $d(d+1)/2$ if $\dim \cH = d$. Indeed, $\vartheta$
is antilinearly Hermitian iff $\langle \psi,\vartheta \varphi\rangle$
is symmetric. With respect to a basis the condition restricts
the off-diagonal entries only. Complex diagonal entries are
allowed.

One calls $\vartheta$ {\em antilinearly unitary} or simply
{\em antiunitary} iff $\vartheta^{\dagger} = \vartheta^{-1}$.
Basic knowledge about antiunitary operators is due to
Wigner \cite{Wigner}.
A conjugation, $\Theta$, is an antiunitary satisfying
$\Theta^2 = \1$.
Writing $\Theta = \Theta^{-1} = \Theta^{\dagger}$ shows
the hermiticity (self-adjointness) of conjugations.
 Well studied
examples are time reversal operators \cite{T}
for Bose particles and for quantum systems with total
integer angular momentum .\\
A conjugation $\Theta$ distinguishes in $\cH$ a real
subspace, $\cH_{\Theta}$, consisting of all $\Theta$-invariant
vectors, i.~e. of all eigenvectors of $\Theta$
with eigenvalue 1. No real subspace in $\cH$ is properly larger
than $\cH_{\Theta}$. Due to Hermiticity,
$\Theta \psi = \psi$ and $\Theta \varphi = \varphi$ result in
$$
\langle \psi, \varphi \rangle =  \langle \varphi, \psi \rangle
$$
so that the scalar product becomes real if restricted
to $\cH_{\Theta}$. In other words, $\cH_{\Theta}$ is not
only a real subspace, it is a {\em real Hilbert} subspace.
On the other hand, $\Theta$  can be gained as complex
conjugation in {\em every} basis contained in $\cH_{\Theta}$.
This establishes a one--to--one
correspondence between maximal real Hilbert subspaces and
conjugations.\\
In a 1-qubit space, i.~e. $\dim \cH =2$, a conjugation
induces a reflection of the Bloch sphere at a certain plane
through its center.
Selecting the 1-2--plane, the plane perpendicular to the 3-axis,
as invariant plane, the effect of the conjugation to the Hermitian
operator
\begin{equation} \label{pauli1}
\varrho =
{1 \over 2}(x_0 \1 + x_1 \sigma_1 + x_2 \sigma_2 + x_3 \sigma_3),
\end{equation}
that is $\varrho \mapsto \tilde \varrho \equiv \Theta \varrho \Theta$,
reads
\begin{equation} \label{pauli2}
\tilde \varrho =
{1 \over 2}(x_0 \1 + x_1 \sigma_1 + x_2 \sigma_2 - x_3 \sigma_3).
\end{equation}

Given a conjugation and a state vector, $\psi$, we shall consider
the absolute value of the transition amplitude between
$\psi$ and $\Theta \psi$, or, what is the same, the square
root of the transition probability between them. The quantity
in question, $|\langle \psi, \Theta \psi\rangle|$, is well
defined for pure states. The problem addressed in the paper
is to extend it to all states in a canonical way.
In other words, we look for functions on the state space
which are completely determined
by their pure state behaviour. This can be
done by relying on the convex nature of the set of all
density operators (states) which reflects the process of
performing Gibbsian mixtures, i.e. of convex sums.
There is one and only one largest
convex function coinciding at pure states with
$|\langle \psi, \Theta \psi\rangle|$, and, following Wootters,
I call it $\Theta${\em --concurrence}. And there is exactly one
smallest concave function within all functions which are
concave extensions
from the chosen values for pure states to all density operators.
That function I call $\Theta${\em --fidelity}.

\section{Fidelity and Concurrence}

Let $\varrho$ and $\omega$ be two density operators in an
Hilbert space $\cH$.
Their {\em transition probability} is
denoted by $P(\varrho, \omega)$, their {\em fidelity}, the square
root of the transition probability, is called
$F(\varrho, \omega)$. It holds
\begin{equation} \label{fidelity}
\sqrt{P(\varrho, \omega)} = F(\varrho, \omega) = {\rm tr} \,
(\sqrt{\omega} \varrho \sqrt{\omega})^{1/2}
\end{equation}
Let $\cH^a$ be an ancillary Hilbert space. For any two vectors,
$\varphi, \psi \in \cH \otimes \cH^a$, which reduce to $\varrho$
and $\omega$, $$ \varrho = {\rm Tr}_a
|\varphi\rangle\langle\varphi|, \quad \omega = {\rm Tr}_a
|\psi\rangle\langle\psi|, $$ the transition amplitude is bounded
from above by the fidelity, $|\langle \varphi, \psi \rangle| \leq
F(\varrho, \omega)$. Indeed, $F(\varrho, \omega)$ is the least
number which fulfills this condition. Equivalently, as $F^2 = P$,
a suitably chosen von Neumann measurement in an ancillary system
can cause a transition $\varrho \mapsto \omega$ with probability
$P(\varrho, \omega)$. A larger transition probability, however, is
not possible \cite{fidel}. The joined concavity of the fidelity
can be seen from
\begin{equation} \label{minf}
F(\varrho, \omega) = \inf_X {1 \over 2} \,
\Bigl( \, {\rm tr}(X \varrho) + {\rm tr}(X^{-1} \omega) \, \Bigr)
\end{equation}
where $X$ runs through all positive and invertible operators $X$.
A proof for finite-dimensional Hilbert spaces is as follows:
Abbreviate by $a$ and $b$ the traces over $X \varrho$ and
$X^{-1} \omega$ respectively. From \cite{inf} one knows
$F^2 \leq ab$. But $2 \sqrt{ab} \leq a+b$, and the right hand
side of (\ref{minf}) cannot be smaller than the left one.
If the
density operators are invertible then there is a unique
positive solution $X$ of
$$
X \varrho X = \omega, \quad X = \varrho^{-1/2}
(\varrho^{1/2} \omega \varrho^{1/2}) \varrho^{-1/2}
$$
With this solution we get $a=b$ and $a=F$, and (\ref{minf}) is
saturated. Now we use continuity to extend the proof to
all pairs of density operators. See also \cite{AU99}.\\
It is useful to extend the
equations (\ref{fidelity}), (\ref{minf}), and similar ones
to all positive operators with finite trace. The simple scaling
properties of $P$, $F$, and related quantities make this is an easy
task. Of course, the physical interpretation of $P$ as a probability
is bound to normalized density operators.

Let $\Theta$ be a conjugation in an Hilbert space $\cH$ and
abbreviate $\tilde \varrho := \Theta \varrho \Theta$.
It is evident from (\ref{minf}) that
\begin{equation} \label{ftheta}
F_{\Theta}(\varrho) := F(\varrho, \tilde \varrho)
\end{equation}
is concave in $\varrho$, \cite{fuchs}. (\ref{ftheta}) will be
called {\em $\Theta$--fidelity of} $\varrho$.\\

In order to introduce the (pre)concurrence \cite{BDSW96}
and \cite{Woo97}
we need the ordered singular numbers, $\lambda_1 \geq \lambda_2
\geq \dots$, of $\sqrt{\varrho} \sqrt{\omega}$, that
is
\begin{equation} \label{spectrum}
\{ \, \lambda_1 \geq \lambda_2 \geq \dots\, \} =
\hbox{spectrum of } \, (\sqrt{\varrho} \omega \sqrt{\varrho})^{1/2}
\end{equation}
Having in mind Wootters' explicit expression for the entanglement
of formation it is tempting to define for any two
density operators (whether normalized or not) the function
\begin{equation} \label{fcon}
C(\varrho, \omega) := \max \{ 0, \lambda_1 - \sum_{j > 1} \lambda_j \}
\end{equation}
and to call it {\em concurrence} of $\varrho$ and $\omega$.\\
A useful relation can be obtained if the rank of $\varrho \omega$
does not exceed two. Adding $P = F^2$ to $C^2$ the cross terms in
the two non-vanishing eigenvalues cancel. But the sum of the squared
eigenvalues (\ref{spectrum}) is equal to the trace of $\varrho \omega$.
Hence
\begin{equation} \label{rank2}
C(\varrho, \omega)^2 + F(\varrho, \omega)^2 =
2 {\rm Tr} (\varrho \omega) \, \hbox{ if rank} \, (\varrho \omega)
\leq 2
\end{equation}
Finally, given a conjugation $\Theta$, we call
{\em $\Theta$--congruence of} $\varrho$ the concurrence
between $\varrho$ and its conjugate $\tilde \varrho$,
\begin{equation} \label{condef}
C_{\Theta}(\varrho) := C(\varrho, \tilde \varrho),
\quad \tilde \varrho = \Theta \varrho \Theta
\end{equation}

In contrast to the higher dimensional cases it is not hard to
get explicit expressions if $\dim \cH = 2$.
With $\varrho$ given by (\ref{pauli1})
and a conjugation acting as in (\ref{pauli2}) one obtains
\begin{equation} \label{dim2}
F_{\Theta}(\varrho) = \sqrt{x_0^2 - x_3^2},
\quad
C_{\Theta}(\varrho) = \sqrt{x_1^2 + x_2^2},
\end{equation}
The next issue is to prove:
$F_{\Theta}$ {\em is a concave and $C_{\Theta}$ is a convex roof
for every conjugation $\Theta$ in every finite dimensional
Hilbert space.} For the time being the finite dimensionality
of the Hilbert space is essential due to some unexamined
mathematical problems in the case of infinite dimensions.
Thus, in all what follows, $\dim \cH = d < \infty$.

\section{Properties of $\Theta$--fidelity and $\Theta$--concurrence}
In this section we derive some implications from and start proving of

{\bf Theorem 1} : \, Let $\Theta$ be a conjugation. Then
\begin{eqnarray}
C_{\Theta}(\varrho) &=&
\min \sum | \langle \phi_k| \Theta |\phi_k\rangle |,
\nonumber\\
F_{\Theta}(\varrho) &=&
\max \sum | \langle \phi_k| \Theta |\phi_k\rangle |,
\label{max-1}
\end{eqnarray}
{\em where the $\min$ and $\max$ has to run through all ensembles
$\{ \phi_1, \phi_2, \dots \}$ such that}
\begin{equation} \label{max-2}
\varrho = \sum |\phi_k\rangle \langle\phi_k|
\end{equation}
{\em is valid.}

The proof of the theorem will terminate in the next section. Up to
that point we consider (\ref{max-1}) as a definition of its
left-hand-sides, and we shall draw conclusions {\em without using}
(\ref{ftheta}) and (\ref{condef}) of the preceding section.

Consider first the case $\varrho = |\psi\rangle\langle\psi|$. Clearly,
every decomposition (\ref{max-2}) is gained by $\phi_k = a_k \psi$
with numbers $a_k$ satisfying $\sum |a_k|^2 = 1$. Hence
\begin{equation} \label{pure}
C_{\Theta}( |\psi\rangle\langle\psi| ) =
F_{\Theta}( |\psi\rangle\langle\psi| ) =
| \langle\psi| \Theta |\psi\rangle |
\end{equation}
A simple consequence of (\ref{max-1}) is homogeneity. For positive reals
\begin{equation} \label{homogen}
C_{\Theta}(\mu \varrho)  =    \mu C_{\Theta}(\varrho), \quad
F_{\Theta}(\mu \varrho)  =    \mu F_{\Theta}(\varrho), \quad
\forall \, \mu \geq 0
\end{equation}
Being in finite dimension the minimum (maximum) in (\ref{max-1})
will be attained by certain decompositions (\ref{max-2}). They
are called {\em optimal decompositions}.\\
Choosing optimal decompositions for $C_{\Theta}(\varrho)$ and
$C_{\Theta}(\omega)$, their union is a decomposition for
$C_{\Theta}(\varrho+\omega)$, though not necessarily an
optimal one. Hence $C_{\Theta}(\varrho) + C_{\Theta}(\omega)$ is
an upper bound for $C_{\Theta}(\varrho + \omega)$. Similar
reasoning can be done for the $\Theta$--fidelity. Thus
\begin{eqnarray}
C_{\Theta}(\varrho + \omega) &\leq&
C_{\Theta}(\varrho) + C_{\Theta}(\omega)
\nonumber\\
F_{\Theta}(\varrho + \omega) &\geq&
F_{\Theta}(\varrho) + F_{\Theta}(\omega)
\label{s-add}
\end{eqnarray}
showing {\em subadditivity of $\Theta$--concurrence} and
{\em superadditivity of $\Theta$--fidelity}. Because of its
homogeneity (\ref{homogen}) we conclude:\\
$C_{\Theta}$ {\em is convex}, $F_{\Theta}$ {\em is concave}.

Now we can go a step further, again without using arguments
from the preceding section. Let $\Omega$ be the state space,
i.~e.~the convex set of normalized density operators. If
$\varrho$ is in this set, a decomposition (\ref{max-2})
can be rewritten as a convex combination
\begin{equation} \label{deco}
\varrho = \sum p_k \pi_k, \quad \pi_k =
{|\phi_k\rangle \langle\phi_k| \over \langle\phi_k | \phi_k\rangle }
\end{equation}
Assuming that our decomposition (\ref{deco}) is optimal for, say,
the $\Theta$--concurrence, we can write
$$
C_{\Theta}(\varrho) = \sum p_k C_{\Theta}(\pi_k)
$$
We
conclude as following \cite{Uh98a}. Let $C'$ be another convex
function on $\Omega$ coinciding with $C$ at the pure states.
Then we have
$$
C'(\varrho) \leq \sum p_k C'(\pi_k) = \sum p_k C_{\Theta}(\pi_k)
$$
But for a an optimal decomposition which of the
$\Theta$--concurrence the right hand sides coincides
with $C_{\Theta}(\varrho)$.
A similar proof is for $F_{\Theta}$. It results

{\bf Theorem 2} : \,
$C_{\Theta}$ {\em is the largest convex function
and} $F_{\Theta}$ {\em is the smallest concave function
on the state space coinciding with}
$|\langle\psi |\Theta| \psi\rangle|$ {\em at the pure states.}

To show that the right hand sides of (\ref{max-1}) coincide
with the definitions used in section 2, optimal decompositions
will be gained in the next section.

\section{Optimal decompositions}
In building optimal decompositions for
our $\Theta$--fidelity and $\Theta$--concurrence the properties of
antilinear operators play a decisive role.  Fix a density operator
$\varrho$ and define an antilinear operator $\vartheta$ by
\begin{equation} \label{antih}
\vartheta \equiv \vartheta_{\varrho} :=
\sqrt{\varrho} \, \Theta \, \sqrt{\varrho}
\end{equation}
Because $\Theta^{\dagger} = \Theta$, $\vartheta$ is antilinearly
Hermitian. Hence
$$
\langle \varphi, \vartheta \psi \rangle = \langle \psi,
\vartheta \varphi \rangle
$$
Substituting $\varphi = \vartheta \psi$
proves all the expectation values of $\vartheta^2$ real and not
negative.  Thus, $\vartheta^2$ is a linear positive operator and the
same is with $\sqrt{\vartheta^2}$.  Let us abbreviate
$\tilde \varrho = \Theta \varrho \Theta$, so that $\vartheta^2$
can be written $\sqrt{\varrho} \tilde \varrho \sqrt{\varrho}$.
Remark, just to see what is going on, how the eigenvalues of
the positive square root of $\vartheta^2$ have been used in
section II to express $F_{\Theta}$ and $C_{\Theta}$.\\
Our next aim is to prove the existence of
a conjugation, $\Theta_0$, depending on $\varrho$, with
which we can polar decompose
\begin{equation} \label{antip}
\vartheta = \Theta_0 \sqrt{\vartheta^2} =
\sqrt{\vartheta^2} \Theta_0, \quad
\vartheta^2 = \sqrt{\varrho} \tilde \varrho \sqrt{\varrho}.
\end{equation}
Let $\lambda^2$, $\lambda > 0$ be an eigenvalue of $\vartheta^2$
and $\cH^{\lambda}$ the Hilbert subspace of the corresponding
eigenvectors. With $\psi$ also $\vartheta \psi$ belongs to
$\cH^{\lambda}$. Define on $\cH^{\lambda}$ the action
$\Theta_0 \psi := \lambda^{-1} \vartheta \psi$. On
$\cH^{\lambda}$ the operator $\Theta_0$ is a conjugation which
commutes with $\vartheta$. If one eigenvalue of $\vartheta^2$
is zero, $\Theta_0$ should induce on $\cH^0$ an arbitrarily
chosen conjugation.
Now $\cH$ is decomposed as a direct orthogonal sum of Hilbert
spaces of the form $\cH^{\lambda}$ and $\Theta_0$ is given
as an operator on every one of them. But this defines $\Theta_0$
uniquely as a conjugation on $\cH$, and (\ref{antip})
is proved. Choosing in every $\cH^{\lambda}$ a
$\Theta_0$-invariant basis, we get a common
eigenbasis, $\{ \psi_1, \psi_2, \dots \}$, such that
\begin{equation} \label{antie}
\vartheta \psi_k = \sqrt{\vartheta^2} \psi_k = \lambda_k
\psi_k, \quad \Theta_0 \psi_k = \psi_k
\end{equation}
with ordered eigenvalues $\lambda_1, \geq \lambda_2, \geq \dots$.

The vectors constituting an optimal decomposition will be obtained
by the help of real Hadamard matrices.
They can be inductively gained by
\begin{equation} \label{hadamard}
A_2 = \pmatrix{1 & 1 \cr 1 & -1 \cr}, \quad
A_{2m} := \pmatrix{A_m & A_m \cr A_m & - A_m \cr}
\end{equation}
for $m = 2, 4, 8, \dots$.
Let us denote by $a_{ki}$ the matrix elements of $A_m$.
These entries are either $1$ or $-1$. They fulfill
\begin{equation} \label{ortho}
\sum_{k=1}^m a_{ki} a_{kj} = m \delta_{ij}, \quad
a_{1 j} = 1 \, \, \forall j
\end{equation}
The number $m$ is adjusted to the dimension $d$ of $\cH$ by
\begin{equation} \label{dim}
m = 2^{n+1}, \quad 2^n < \dim \cH \leq 2^{n+1}
\end{equation}
With an arbitrary selection of $d$ unimodular numbers
(phase factors), $\epsilon_1, \epsilon_2, \dots$, we define
with a basis (\ref{antie}) the vectors
\begin{equation} \label{optdeco1}
\varphi_k =
\sum_{i=1}^d a_{ki} \epsilon_i \psi_i,
\quad k=1, 2, \dots, m
\end{equation}
By the help of (\ref{optdeco1}) and (\ref{ortho}) it
is straightforward to prove the following, essentially known
identities
\begin{eqnarray}
\sum_{k=1}^m | \varphi_k \rangle\langle \varphi_k | &=&
m \, \sum_{i=1}^d | \psi_i \rangle\langle \psi_i | = m \, \1
\nonumber\\
\langle \varphi_k | \vartheta | \varphi_k \rangle
&=&  \, \sum_{j=1}^d \epsilon_j^{-2} \lambda_j
\label{optdeco2}
\end{eqnarray}
The remarkable deviation from most uses of Hadamard matrices
is in the appearance of the phase factors produced by
the antilinearity of $\vartheta$. They provide sufficient
flexibility in adjusting the expectation values of $\vartheta$.
By varying the $\epsilon_j$
in the second equation arbitrarily, the absolute values of the
numbers $\langle\varphi_k| \vartheta |\varphi_k\rangle$
fill completely the following interval of real numbers:
\begin{equation} \label{minmax}
\sum_{j=1}^d \lambda_j \geq |\sum_{j=1}^d \epsilon_j^{-2} \lambda_j|
\geq \max \{ 0, \, \lambda_1 - \sum_{j=2}^d \lambda_j \}
\end{equation}
Proof: a) The sum of the $\lambda_j$ is an upper bound (triangle
inequality) and it is reached with
$\epsilon_j^{-2} = 1$ for all $j$. The simplest choice is
$\epsilon_j = 1$ for all $j$. b) If the $\lambda_1$
is not smaller than the sum of the remaining lambdas, a lower
bound is $|\epsilon_1^{-2} \lambda_1 - x|$ where $x$ is the
maximum absolute value of
$\epsilon_2^{-2} \lambda_2 + \epsilon_3^{-2} \lambda_3 + \dots$.
Hence we get the asserted lower bound. The bound is
attained for $\epsilon^{-2} = 1$ and $\epsilon_j^{-1} = i$ for
$j>1$. c) It remains to prove: If the assumption of b) is not
valid, the lower bound 0 should be reachable. In this case
$$
\sum_{j=2}^d \lambda_j > \lambda_1 >
\lambda_2 - \sum_{j=3}^d \lambda_j
$$
The first inequality is the assumption, the second follows because
otherwise $\lambda_1 < \lambda_2$ in contradiction to the assumed
ordering of the $\lambda_k$.
We like to conclude the existence of a representation
$$
\lambda_1 = | \sum_{j=2}^d \epsilon_j^{-2} \lambda_j |
$$
as then the lower bound zero can be reached: We have to prove
the same assertion as above, but now the length of the sum
is $d-1$. Hence the
proof is done if (\ref{minmax}) is true for sums of length less
than $d$. Starting with $d=2$, the proof terminates by induction
to the length of the sum to be estimated.

Given $\lambda_1, \lambda_2, \dots$ we choose unimodular numbers
$\epsilon_1, \epsilon_2, \dots$ saturating respectively the upper
bound or the lower bound of (\ref{minmax}). With this
choice the vectors (\ref{optdeco1}) are denoted by $\varphi^+_k$
(to refer to the upper bound) and by $\varphi^-_k$
(to indicate the use of the lower bound) respectively.
From the construction follows that the insertion of
\begin{equation} \label{optdeco3}
\phi^-_k = \sqrt{\varrho} \varphi^-_k, \quad
\phi^+_k = \sqrt{\varrho} \varphi^+_k, \quad
k = 1, \dots, m
\end{equation}
into (\ref{max-2}) estimates (\ref{max-1}) as follows:
$$
C_{\Theta} \leq \max \{ 0, \, \lambda_1 - \sum_{j=2}^d \lambda_j \},
\quad F_{\Theta} \geq \sum \lambda_j
$$
These inequalities must be equalities. For the proof
we use an arbitrary
decomposition $\1 = \sum |\chi_k\rangle\langle\chi_k|$
of the unity, insert
$\phi_k = \sqrt{\varrho} \chi_k$ into (\ref{max-1}),
and convert the sum to be estimated by the help of
(\ref{antie}) into
$$
a \equiv \sum | \langle \phi_k, \Theta \phi_k\rangle | =
\sum_k \sum_{j=1}^d | \lambda_j  ( \langle \chi_k, \psi_j )^2 |
$$
At first we estimate concurrence by choosing the phases of
$\psi_j$ such that $\langle \chi_1, \psi_k \rangle$
becomes real and positive. We get
$$
a = | \lambda_1 - b |, \quad b \equiv
\sum_{j=2}^d \lambda_j \sum_k ( \langle \phi_k, \psi_j )^2 |
$$
for the sum in question. If $|b|$ is larger than $\lambda_1$
we already obtained $a=0$ with $\phi_k = \phi^-$. In the other
case $|b|$ cannot exceed $\lambda_2 + \lambda_3 + \dots$, i.e.
$a - |b| \leq C_{\Theta}$.\\
Concerning the $\Theta$--fidelity the Schwarz inequality will be applied
to the positive Hermitian form $\langle \phi, \sqrt{\vartheta^2}
\phi'\rangle$.  Respecting (\ref{antip}) and (\ref{antie}) one gets $$ |
\langle \phi_k, \sqrt{\vartheta^2} \Theta_0 \phi_k\rangle | \leq \langle
\phi_k, \sqrt{\vartheta} \phi_k\rangle $$ Therefore $a$ cannot be larger
than the trace of $\sqrt{\vartheta}$.  The latter is equal to
$F_{\Theta}$ and we arrive at $a \leq F_{\Theta}$. $\Box$\\
We have not only proved theorem 1 but also

{\bf Corollary 3 :} \, {\em
Let $\dim \cH = d$ and $2^n < d \leq 2^{n+1}$.
For every $\varrho$ there exist optimal decompositions for the
$\Theta$--concurrence the length of which does not exceed $2^{n+1}$.
The same is true for the $\Theta$--fidelity.}\\
Remarks: a) Can the bounds for the optimal length become more
stringent for certain dimensions of $\dim \cH$.  The construction
above seems to deny it. But a proof is missing.
b) If $d = 4 = 2 \times 2$, then
$n=2$ and there are optimal decompositions of maximal
length four as shown by Wootters.
See also \cite{length} for the optimal length problem.

\section{Roofs}
We now call attention to some peculiarities of convex or
concave function on the state space which admit optimal
decompositions. These functions are quite different from
unitarily invariant ones like, for instance, von Neumann
entropy. The latter do not at all discriminate between pure
states, they just estimate how strongly a state is mixed.
Roofs, as defined below, and in particular convex or concave
ones, draw all their information from their values at pure
states. They try to interpolate between those values as
linearly as possible. Let us see how it is achieved by
two  simple examples.

In two dimensions $\Theta$-fidelity and $\Theta$-concurrence
are given by $\sqrt{1 - x_3^2}$ and $\sqrt{x_1^2 + x_2^2}$
on the unit ball $x_1^2+x_2^2+x_3^2 \leq 1$, see (\ref{dim2}).
The first one remains constant on the
planes $x_3 =$ constant, the second one does so along the lines
$x_1 = c_1$, $x_2 = c_2$. The intersections of a plane or of a
straight line with the unit ball are not only convex:
The intersections can be gained as the convex hulls of
the pure states they contain.

In turning to the general case we denote by $\Omega$ the convex
set of all normalized density
operators on a finite dimensional Hilbert space and by
$\Omega^{\rm pure}$ the set of its extremal points, i.~e.~the
set of pure density operators.\\
A convex subset $\Omega_0$ of $\Omega$ will be called a
{\em convex leaf of} $\Omega$ iff
\begin{equation} \label{leaf1}
\Omega_0 = \hbox{convex hull of} \, (\Omega_0 \cap
\Omega^{\rm pure})
\end{equation}
Let $G = G(\varrho)$ be a function on $\Omega$ and $\Omega_0$
a convex leaf of $\Omega$. $G$ is called {\em convexly linear}
(or, equivalently, {\em affine} or {\em flat}) on $\Omega_0$ if
for all probability vectors $p_1, p_2, \dots$ and for all
choices of pure states
\begin{equation} \label{leaf3}
\pi_1, \pi_2, \dots \, \in \Omega_0 \cap \Omega^{\rm pure}
\end{equation}
$G$ satisfies the relation
\begin{equation} \label{leaf4}
G(\sum p_j \pi_j) = \sum p_j G(\pi_j)
\end{equation}
It is not necessary to check condition (\ref{leaf4}) for all possible
convex linear combinations in case $G$ is either convex
or concave:

{\bf Lemma R-1:} \, {\em Let $G$ be convex or concave. If
\begin{equation} \label{d1}
\varrho = \sum q_j \pi_j, \quad q_k > 0
\end{equation}
is a decomposition of $\varrho$ into pure density operators
$\pi_1, \pi_2, \dots$, and if
\begin{equation} \label{d2}
G(\varrho) = \sum q_j G(\pi_j)
\end{equation}
 is valid then $G$ is convexly linear
on the convex hull of $\pi_1, \pi_2, \dots$.}\\

Proof: Assume $G$ is convex. Given $\varrho$, there is convexly
linear function $l$ satisfying $G \geq l$ on $\Omega$, and
$G(\varrho) = l(\varrho)$. Together with (\ref{d2}) we get
$$
l(\varrho) = G(\varrho) = \sum q_j G(\pi_j) \geq
\sum q_j l(\pi_j)
$$
Because the right hand term is $l(\varrho)$, the $\geq$
symbol must be an equality sign. But $G \geq l$ now enforces
$l(\pi_j) = G(\pi_j)$ for the pure states involved in (\ref{d2}).
By the help of this equalities we estimate $G(\omega)$,
$\omega = p_1 \pi_1 + \dots$, by
$$
l(\omega) \leq G(\omega) \leq \sum p_k G(\pi_k) = l(\omega)
$$
and the inequality must be an equality.
(The first inequality sign is due to $l \leq G$, the second due
to the convexity of $G$.) This proves the lemma for convex $G$.
Because $-G$ is convex if $G$ is concave, the
lemma remains true for concave functions.
Another proof is in \cite{BNU96}.

By definition, $G$ is a {\em roof} if $\Omega$ can be covered by
convex leaves such that $G$ is convexly linear on every leaf of
the covering. The covering is said to be a {\em convex covering
belonging to} or {\em compatible with $G$.}

There is a simple geometric picture beyond. Assume a real
number $g=g(\pi)$ is given for every pure state $\pi$.  The
idea is to think of a wall, made of straight lines starting
from $\pi$ and terminating at $g(\pi) \pi$. The demand is,
to cover the state space $\Omega$ by a roof, founded upon the
wall, which is as flat as possible.
To satisfy the demand one joins every two points
on the wall by a straight line,
every three points by a triangle, and so on.
If the dimension of the polyhedra becomes large enough,
$(\dim \cH)^2$ in our case, the set of polyhedra covers
$\Omega$, (an application of Caratheodory's theorem),
and we stop. To get a roof we have to select a {\em one-fold}
covering of $\Omega$ from our huge set of polyhedra: There
should be a function $\omega \to G(\omega)$
such that $x=G(\omega)$  whenever $x \omega$ is contained in
one of the polyhedra of the selected covering. If it occurs,
$G(\omega) \omega$ is a convex combination of the
$g(\pi_j) \pi_j$ which generate the polyhedron. Taking the trace
yields a representation
$$
G(\omega) = \sum_{j=1}^m p_j g(\pi_j), \quad m \leq (\dim \cH)^2
$$
From the bewildering manifold of roofs we select the highest
(or the lowest) one: Given $\omega$ we look for a polyhedron
containing $x \omega$ with the largest (or with the smallest)
possible real number $x$. Let us call this number $G^+(\omega)$
respectively $G^-(\omega)$. There is such a polyedron if $g$
is continuous, because then the set of
all polyhedra based on a bounded number of edges is compact.\\
Some generalities can be abstracted from the construction
above, see \cite{Uh97a}, \cite{Uh98a}. They are summarized
in the following lemma.

{\bf Lemma R-2 :} \, {\em Let $g = g(\pi)$ be a real
and continuous function on the set of pure states.\\
a) \, There is exactly one convex roof
$G^-$ and exactly one concave roof $G^+$ on $\Omega$
which coincides on $\Omega^{\rm pure}$ with $g$.\\
b) \, $G^+$ is the smallest concave function and $G^-$ is the
largest convex function which coincides at the pure states with
$g$.\\
c) It is
$$
G^+(\varrho) = \max \sum p_j g(\pi_j)
$$
$$
G^-(\varrho) = \min \sum p_j g(\pi_j)
$$
where the variations  have to run through all convex decompositions of
$\varrho$ with pure states.}

Starting the discussion above from
$g(\pi) = |\langle \psi, \Theta \psi\rangle|$, where
$\pi = |\psi\rangle\langle\psi|$ and $\Theta$ is a conjugation,
we arrive at $G^+ = F_{\Theta}$ and $G^- = C_{\Theta}$.
Within the pure states belonging to one of the
optimal decompositions of the preceding section the values $g(\pi)$
remain constant. Hence $G^+$ and $G^-$ are constant on the
convex leaf they generate:

{\bf Corollary 4 :} \, {\em
The $\Theta$--concurrence (respectively the $\Theta$--fidelity)
allows for a convex foliation such that $C_{\Theta}$
(respectively $F_{\Theta}$) is constant over every of its leaves.}

As an immediate consequence, $\varrho \to f(C_{\Theta}(\varrho))$
and
$\varrho \to f(F_{\Theta}(\varrho))$ are roofs over $\Omega$
for every function $f(x)$ defined on the unit interval.
In general the roofs so obtained cease
to be convex or concave. But there are some rules
guaranteeing convexity (concavity) in some cases.
To preserve convexity it suffices that $f$ is convex and increasing.
Concavity is guaranteed with $f$ concave and decreasing \cite{Roc70}.

Examples are: $C_{\Theta}^s$ with
real $1 < s$ is a convex roof, $F_{\Theta}^s$ with
$0 < s < 1$ is a concave roof.
An important convex and  increasing
function, used by Hill and Wootters in
\cite{HW97} and \cite{Woo97} to get an expression
for the entanglement of formation, \cite{BDSW96}, reads
\begin{equation} \label{eofa1}
f_{HW}(x) =: s({1 + \sqrt{1-x^2} \over 2}) +
s({1 - \sqrt{1-x^2} \over 2}),
\end{equation}
where $s(y)$ abbreviates $-y \ln y$. Thus
\begin{equation} \label{eofa2}
\varrho \mapsto f_{HW}( C_{\Theta}(\varrho) )
\end{equation}
is a convex roof for every conjugation in every Hilbert space.
However, {\em only} if the Hilbert space
is 4-dimensional, and $\Theta$ the Hill--Wootters conjugation,
(\ref{eofa2}) is equal to the entanglement of formation.
In bipartite $2 \times 2n$ systems (\ref{eofa2}) can only be a
lower bound to the entanglement of formation
for appropriately chosen $\Theta$. (See the next section.)

The following statements copy facts known in 2-qubit systems to
a more general frame.\\
The maximum of $F_{\Theta}$ is one, and $F_{\Theta}=1$ is
the equation of a convex leaf for $F_{\Theta}$ by lemma R-2.
If $F_{\Theta}(\varrho)=1$
then $\Theta \varrho \Theta = \varrho$ by (\ref{ftheta}),
and $\varrho$ has a basis
of $\Theta$-invariant eigenvectors.\\
The minimum of $C_{\Theta}$ is zero. The set of all states with
vanishing $\Theta$-concurrence is a convex leaf with respect to
$C_{\Theta}$. If $\varrho$ is $\Theta$-invariant,
$\Theta \varrho \Theta = \varrho$, then $C_{\Theta}(\varrho)=0$
if and only if no eigenvalue of $\varrho$ exceeds $1/2$.

The entanglement of formation vanishes, as known from \cite{BDSW96},
exactly for separable,
i.~e. classically correlated states \cite{We89}, \cite{sep}.
Separability in a 2-qubit-system can equally well be characterized
by the vanishing of $C_{\Theta}$, $\Theta$ the Hill--Wootters
conjugation. Again, just for 2-qubits, $F_{\Theta}=1$ is
the equation for the convex hull of the maximally entangled
pure states.

\section{Examples}
This section considers some possible applications of the
general theorems. By looking at examples we ask whether
$\Theta$-concurrences can be used to decide
separability problems in bi-- and multipartite systems.
In a 2-qubit system a density operator is separable if and
only if its concurrence vanishes. Could one suppose similar
statements in a higher dimensional or in a multi-qubit system?
Certainly not with just one
functional. But with sufficiently many it can work. Before
treating the examples we have to return to a further issue in
antilinearity.

All conjugations of an Hilbert space are unitarily equivalent.
From (\ref{spectrum}) and the definitions of fidelity and
concurrence one gets
$$
F_{\Theta'}(\varrho) = F_{\Theta}(U \varrho U^{\dagger}),
\quad
C_{\Theta'}(\varrho) = C_{\Theta}(U \varrho U^{\dagger})
$$
with $\Theta' = U^{\dagger} \Theta U$ and every unitary operator
$U$. However, in a bi-- or multipartite system,
\begin{equation} \label{mpc0}
\cH = \cH^a \otimes \cH^b \otimes \otimes \cH^c \dots ,
\end{equation}
one considers two conjugations equivalent  iff there is
a {\em local} unitary $U$ such that
$\Theta' = U^{\dagger} \Theta U$. Some of these equivalence
classes consist of tensor products of antiunitary operators,
\begin{equation} \label{mpc1}
\Theta = \theta_a \otimes \theta_b \otimes \dots
\end{equation}
To obtain a conjugation, the square of each factor must be
a multiple of the appropriate identity, for example
$\theta_a^2 = c_a \1^a$. According to Wigner there are only
two possibilities, $c_a = \pm 1$. Therefore, a factor in
(\ref{mpc1}) is either a conjugation or it is an antiunitary
satisfying $\theta^2 = -\1$. The number of the latter cases
must be even to obtain a conjugation by (\ref{mpc1}).

For the purpose of the present paper an antiunitary $\theta$
satisfying $\theta^2 = -\1$ is called a {\em skew conjugation}.
While skew conjugations are mostly discussed in connection with
time reversal of fermions, we need them as building blocks
for conjugations in multipartite quantum systems.\\
A skew conjugation fulfills $\theta^{-1} = - \theta^{\dagger}$
and
\begin{equation} \label{mpc2}
\langle \phi, \theta \phi'\rangle +
\langle \phi', \theta \phi \rangle = 0
\end{equation}
All expectation values of a skew conjugation vanish.
There is a consequence for vectors $\psi \in \cH$  which are
separable with respect to the first factor in (\ref{mpc0}),
say $\psi = \phi^a \otimes \varphi$.
If the first antiunitary, $\theta_a$, is a skew conjugation,
the expectation value $\langle \psi, \Theta \psi\rangle$ must
vanish. In other words: Let $\Theta$ be a conjugation
(\ref{mpc1}) and assume its first factor is a skew conjugation.
If $\langle \psi, \Theta \psi \rangle$ is not zero,
$|\psi\rangle\langle\psi|$ cannot be $\cH^a$--separable.

A skew conjugation, $\theta$, allows for a representation \cite{Wigner}
\begin{equation} \label{fc}
\theta \psi_{2j} = \psi_{2j-1}, \quad \theta \psi_{2j-1} =
- \psi_{2j},
\end{equation}
$1 \leq j \leq n$, with a certain basis, $\psi_1, \psi_2, \dots$,
called a $\theta$--basis. By (\ref{fc}) the Hilbert space
decomposes into a direct sum of 2-dimensional,
$\theta$--invariant Hilbert subspaces. Of course, any basis of
$\cH$ can serve as a $\theta$--basis for
a certain skew conjugation $\theta$.

In 1-qubit spaces there
is, up to a phase, just one skew conjugation $\theta$  that
may be defined by $|0\rangle \to i |1\rangle$,
$|1\rangle \to -i |0\rangle$. (The imaginary unit
in the definition is by convention.)
On the state space it induces the well known {\em spin flip}.
With that definition $\theta \otimes \theta$
is the Hill--Wootters conjugation of a 2-qubit space.\\

Example 1: \,
Consider in (\ref{mpc0}) a direct product
$\cH = \cH^a \otimes \cH^b$ of
two even--dimensional Hilbert spaces. We distinguish a
special class, ${\cal F}$, of conjugations: $\Theta \in {\cal F}$
if the conjugation can be written as the product
$\Theta = \theta^a \otimes \theta^b$ of two skew
conjugations. Notice that, up to a phase, ${\cal F}$
consists of one conjugation in the 2-qubit case,
the Hill--Wootters one.\\
We have already seen from (\ref{mpc2})
that for this class  $\langle \psi| \Theta |\psi\rangle =0$ if
$\psi$ is a product vector. Thus $C_{\Theta}(\pi) = 0$ for
every pure product state and for every $\Theta \in {\cal F}$.
But, as seen at the end of the preceding section, the equation
$C_{\Theta}(\varrho)=0$ defines a convex leaf, i.~e.
{\em $C_{\Theta}$ vanishes for all separable density operators}.
One may rephrase the statement by saying: {\em If $\varrho$ is a
state in a bipartite system and
if we can find $\Theta \in {\cal F}$ such that $C_{\Theta} > 0$,
then $\varrho$ cannot be separable.}\\
We now complement the last statement: {\em Let $\pi$ be pure.
If $C_{\Theta}(\pi)=0$ is true for all $\Theta \in {\cal F}$
then $\pi$ is a product state, i.e. separable.}\\
For the proof we consider an arbitrary unit vector $\psi \in \cH$
and assume $\dim \cH^a = 2n \leq \dim \cH^b$. We use the Schmidt
decomposition
\begin{equation} \label{schmidt}
\psi = \sum \alpha_j \phi^a_j \otimes \phi^b_j, \quad
\alpha_1 \geq \alpha_2 \geq \dots
\end{equation}
to define a skew conjugations in the two parts of our
bipartite system. $\theta_a$ is defined by requiring
$\phi^a_1, \phi^a_2, \dots$ to be a $\theta_a$--basis. In
$\cH^b$ we complete, if necessary, the $\phi^b_j$ vectors to
a basis which then is used as a the defining
$\theta_b$--basis. After these preparations we consider
$\Theta = \theta_a \otimes \theta_b$, a conjugation tailored for
the vector (\ref{schmidt}). A straightforward calculation
yields
\begin{equation} \label{cschmidt}
\langle\psi| \Theta |\psi\rangle =
2 \sum_{j=1}^n \alpha_{2j} \alpha_{2j-1}
\end{equation}
The sum on the right-hand-side can vanish only if all the
Schmidt coefficients $\alpha_j$ vanish with the exception of
the largest one. Hence $\psi$ must be a product state.

Can we skip in the last statement the purity requirement?
It seems unlikely with the exception of the 2-qubit case.
Thus we are faced with the problem to characterize the set of states
with vanishing $\Theta$-concurrences for
all conjugations from ${\cal F}$. Let us call the set of all
these states $\Omega^c$. As an intersection of convex leaves
it is convex, but not necessarily a leaf. It contains all
separable states. Moreover, a pure state is in $\Omega^c$ if
and only if it is separable. But not all extremal
points of $\Omega^c$ might be pure and, then, it will
contain density operators which are not separable.\\

Example 2: \,
We proceed with the setting of example 1 and require
$\cH^a$ to be 2-dimensional. The requirement allows to bound
the entanglement of formation from below for any even
dimensional second factor ${\cH^b}$ in the bipartite system.
To do so we use (\ref{cschmidt}) to establish
\begin{equation} \label{cschmidt1}
2 \sqrt{\det \rho} =
\sup_{\Theta} | \langle \psi, \Theta \psi \rangle|,
\quad \Theta \in {\cal F},
\end{equation}
Here $\rho$ denotes the partial trace of
$|\psi \rangle\langle \psi\rangle|$ over the second factor,
$\cH^b$. It then follows a lower bound for the entanglement
of formation, $E(\varrho)$.
\begin{equation} \label{eofa3}
E(\varrho) \geq \sup_{\Theta} f_{HW}( C_{\Theta}(\varrho) )
= f_{HW}( \sup_{\Theta} C_{\Theta}(\varrho))
\end{equation}
$f_{HW}$ is explained by (\ref{eofa1}). The equality sign is
due to the monotonicity of $f_{HW}$.
The right hand side of (\ref{eofa3}) is convex as a
sup of convex functions of type (\ref{eofa2}). For pure
states it coincides by (\ref{cschmidt1}) with the entanglement
of formation. But the entanglement of formation is a convex roof
by its definition, see \cite{BDSW96} and point c) of lemma R-2.
Hence the left hand side is the largest possible convex function
with the described values for pure states.\\

Example 3: \,
Now we try a similar procedure as in example 1 for
a 3-qubit-system. As already mentioned there is,
after fixing a phase, only one skew
conjugation, say $\theta$, in a 2-dimensional Hilbert space.
Every conjugation in dimension two is of the form $U \theta$
with unitary $U$.\\
$\cH$ in (\ref{mpc0}) is now the direct product of three
2-dimensional Hilbert spaces. Consider the  conjugations
\begin{equation} \label{3sep}
U \theta \otimes \theta \otimes \theta, \quad
\theta \otimes U \theta \otimes \theta, \quad
\theta \otimes \theta \otimes U \theta
\end{equation}
Let $\psi \in \cH$ and $\Theta$ from this set. Then
$\langle\psi| \Theta |\psi\rangle$ is zero if $\psi$ is a
product vector. A separable $\varrho$ allows
for a convex decomposition with product states by definition.
For $C_{\Theta}=0$ determines a convex leave,
$C_{\Theta}(\varrho)$ has to vanish.\\
Turn now to the reverse and let be
$\pi$ a pure states with $C_{\Theta}(\pi)=0$ for some
conjugations listed in (\ref{3sep}).
The manifold of pure product states is 8--dimensional. We shall
prove that eight equations $C_{\Theta}=0$ with conjugations from
(\ref{3sep}) are sufficient
to decide whether $\psi$ is a product vector or not.\\
This goes as follows.
Write $\psi$ as a sum
$|0\rangle |\varphi_0\rangle + |1\rangle |\varphi_1\rangle$ and
start by the first set of conjugations listed in (\ref{3sep}).
We have to solve the equations
$$
0 = \langle \psi|\psi\rangle = \sum \langle i|U \theta |j\rangle
\langle \varphi_i | \tilde \varphi_j \rangle
$$
The tilde abbreviates the Hill--Wootters conjugation
$\theta \otimes \theta$.
With unitaries of the form
$U |j\rangle = \epsilon_j |j\rangle$ we see that
$\varphi_k$ is orthogonal to $\tilde \varphi_k$. Hence
$\varphi_k$ is a product vector. To come to this conclusion, we
need two diagonal unitaries. Next, with $U$ equal to either
$\sigma_1$ or $\sigma_2$, we see that $\varphi_0$ is orthogonal to
$\tilde \varphi_1$. Because both are product vectors, either the
first or the second one of their constituents has to be orthogonal
one to another. Hence, after checking $C_{\Theta}=0$ with
4 conjugations from our
list, we arrive, up to a local unitary, at one of two possibilities:
$$
|0\rangle |\phi\rangle |0\rangle + |1\rangle |\phi'\rangle |1\rangle,
\quad
|0\rangle |0\rangle |\phi\rangle + |1\rangle |1\rangle |\phi'\rangle
$$
Choosing now a conjugation from the second group of (\ref{3sep})
yields $\langle \phi| U \theta |\phi'\rangle=0$. We need just two
of them to see that either $\phi=0$ or $\phi'=0$ has to take place,
provided $\phi$ is located at the second position in the direct
product. To cover also the case with $\phi$ in the third position,
we need two conjugations from the third group.\\
{\em Let $\pi$ be a pure state of a 3-qubit system. There are
8 conjugations of the form (\ref{3sep}) such that $\pi$
is a product state if and only if
$C_{\theta}(\pi)=0$ is valid for all of them.}\\
It is tempting to ask whether one can prove similar
statements for any multi-qubit system. I believe the answer is
affirmative, but I did not check it.\\

Last not least we are going to cure a curious shortcoming of
the treatment in example 1: It cannot be applied if one of the factors
of the bipartite system is odd dimensional: The set ${\cal F}$
becomes empty. The same unsatisfactory event arises if no or only one
factor of a multipartite system is even dimensional.\\
Let us think, for example, the factor $\cH^a$ is 3-dimensional.
To get an appropriate antilinear operator $\theta_a$ we split
$\cH^a$ into a direct sum of a 2-dimensional and a 1-dimensional
Hilbert space. In the former we equal $\theta_a$ to a skew
conjugation. In the latter we set $\theta_a$ to zero. We do not
get an antiunitary operator, but an antilinear operator satisfying
$\theta_a^{\dagger} = - \theta_a$. This relation suffices
to guarantee (\ref{mpc2}). It seems natural, therefore, to allow
in (\ref{mpc1}) the larger class of antilinear $\theta$
fulfilling $\theta^{\dagger} = \pm \theta$ as factors, and to
require for the tensor product $\Theta^{\dagger} = \Theta$ only.\\
Returning to the bipartite system of example 1 we could
consider the larger class of antilinear operators
$$
\Theta =\theta_a \otimes \theta_b, \quad \theta_a^{\dagger} = - \theta_a,
\, \, \theta_b^{\dagger} = - \theta_b
$$
so that $\Theta$ is antilinearly Hermitian, and, nevertheless,
$\langle\psi, \Theta \psi\rangle = 0$ for product vectors $\psi$.

We arrive at the following general question: Do $\Theta$-fidelity
(\ref{ftheta}) and $\Theta$-concurrence (\ref{condef}) remain
concave respectively convex roofs for any antilinear
self-adjoint $\Theta$.
Going through all the proofs one finds it essential that the
antilinear operator $\vartheta := \sqrt{\varrho} \Theta \sqrt{\varrho}$
is antilinearly Hermitian. For that reason one proves by literally the
same arguments:

{\bf Theorem 5 :} \, {\em Let $\Theta$ be antilinear and
self-adjoint, $\Theta = \Theta^{\dagger}$. Then
$$
F_{\Theta} := F(\varrho, \Theta \varrho \Theta),
\quad
C_{\Theta} := C(\varrho, \Theta \varrho \Theta)
$$
is a concave respectively a convex roof. Theorem 1 and
Corollaries 3 and 4 remain valid for them.}\\

Example 4: \,
The final aim of the exercise is to determine fidelity
and concurrence of certain
conjugated states of rank two in in a 2-qubit space.
The reader should consider the example as representative
for a lot of others which need more calculation effort.\\
Let $\cH_2$ be a 2-dimensional Hilbert space. The transition
probability can be given by elementary algebraic operations
\cite{Hu92b}. For the present purpose an adequate expression
reads
\begin{equation} \label{P}
P(\varrho, \omega) \equiv F(\varrho, \omega)^2
= {\rm Tr} \varrho \omega + 2 \sqrt{\det \varrho \det \omega}
\end{equation}
By the aid of (\ref{rank2}) the equation can be converted to
\begin{equation} \label{C}
C(\varrho, \omega)^2 = {\rm Tr} \varrho \omega -
2 \sqrt{\det \varrho \det \omega}
\end{equation}
Let $\Theta$ be an antilinear Hermitian operator acting on
$\cH_2$. To get $F_{\Theta}(\varrho)$ or $C_{\Theta}(\varrho)$
we have to know the trace of $\varrho \Theta \varrho \Theta$
and the determinants of $\varrho$ and $\Theta \varrho \Theta$.

After these preliminaries we think of $\cH_2$ as of a subspace
of a 2-qubit Hilbert space $\cH$.
We cannot use the Hill--Wootters conjugation $\Theta_{HW}$ in
(\ref{P}) or (\ref{C}) directly because, generally,  $\cH_2$
will not allow $\Theta_{HW}$ as a symmetry. Therefore we set
$\Theta := Q \Theta_{HW} Q$ with $Q$ the projection operator
projecting $\cH$ onto $\cH_2$. $\Theta$, so defined, will be
antilinearly Hermitian and it maps $\cH_2$ into $\cH_2$.
By the little trick we see, abbreviating
$\tilde \varrho = \Theta_{HW} \varrho \Theta_{HW}$,
$$
F(\varrho, \tilde \varrho ) = F_{\Theta}(\varrho),
\quad
C(\varrho, \tilde \varrho ) = C_{\Theta}(\varrho)
$$
{\em whenever $\varrho$ is supported by $\cH_2$.}

We assume $\cH_2$ is generated by two separable unit vectors,
\begin{equation} \label{e4.1}
\psi_i = \phi^a_i \otimes \phi^b_i \in \cH = \cH^a \otimes \cH^b
\end{equation}
We choose their phases such that
\begin{equation} \label{e4.2}
 \langle \phi^a_0, \phi^a_1 \rangle = a, \quad
 \langle \phi^b_0, \phi^b_1 \rangle = b
\end{equation}
where $a$, $b$ are positive real numbers between 0 and 1. We
get, by appropriately adjusting the free phase in the
Hill-Wootters conjugation,
\begin{equation} \label{e4.3}
\langle \psi_1, \Theta_{HW} \psi_0 \rangle =
\langle \psi_0, \Theta_{HW} \psi_1 \rangle =
\sqrt{(1-a^2)(1-b^2)}
\end{equation}
We can replace $\Theta_{HW}$ by $\Theta = Q \Theta_{HW} Q$
in (\ref{e4.3}) without changing its validity. Remind also that
$\langle \psi_i, \Theta \psi_i \rangle=0$ because
$\psi_i$ is a product vector.\\
We introduce a suitable basis by
\begin{equation} \label{e4.4}
\varphi^+ = {\psi_0 + \psi_1 \over \sqrt{2(1 + ab)}}, \quad
\varphi^- = {\psi_0 - \psi_1 \over \sqrt{2(1 - ab)}}
\end{equation}
By a short calculation one concludes from (\ref{e4.3})
\begin{equation} \label{e4.5}
\Theta \, \varphi^{\pm} = a_{\pm} \varphi^{\pm}, \quad
a_{\pm} = {\sqrt{(1 - a^2)(1 - b^2)} \over 1 \pm ab}
\end{equation}
Possessing a distinguished basis (\ref{e4.4}) in $\cH_2$ we represent
any density operator $\varrho$ supported by $\cH_2$ as usual
by the help of Pauli operators, see (\ref{pauli1}).
The Pauli operators to the basis (\ref{e4.5}) are by convention
$$
\sigma_3 = |\varphi^+ \rangle\langle \varphi^+|  -
|\varphi^- \rangle\langle \varphi^-|, \sigma_1 =
|\varphi^+ \rangle\langle \varphi^-| + |\varphi^- \rangle\langle \varphi^+|
$$
and $\sigma_2 = i \sigma_1 \sigma_3$. Transforming $\varrho$
according to $\varrho \to \Theta \varrho \Theta$ can
be accomplished by transforming the identity
of $\cH_2$ and the just introduced Pauli operators.
Using (\ref{e4.5}),
\begin{eqnarray}
\Theta^2 = \Theta Q \Theta  &=&
{a_+^2 + a_-^2 \over 2} \1 + {a_+^2 - a_-^2 \over 2} \sigma_3
\nonumber\\
\Theta \sigma_3 \Theta &=&
{a_+^2 - a_-^2 \over 2} \1 + {a_+^2 + a_-^2 \over 2} \sigma_3
\nonumber
\end{eqnarray}
$$
\Theta \sigma_1 \Theta = a_+ a_- \sigma_1,
\quad
\Theta \sigma_2 \Theta = - a_+ a_- \sigma_2
$$
One gets for the determinant
$$
\det \Theta \varrho \Theta = (a_+ a_-)^2 \det \varrho
$$
and for the trace of $\varrho \Theta \varrho \Theta$
$$
{a_+^2 + a_-^2 \over 4} (x_0^2 + x_3^2) + {a_+^2 - a_-^2 \over 2}
x_0 x_3 + {a_+ a_- \over 2}(x_1^2 - x_2^2)
$$
These expressions shall be inserted into (\ref{P}) and (\ref{C}).
\begin{eqnarray}
F_{\Theta}(\varrho)^2 &=& {1 \over 4} [(a_+ + a_-) x_0
+ (a_+ - a_-) x_3 ]^2 -  a_+ a_- x_2^2,
\nonumber\\
C_{\Theta}(\varrho)^2 &=& {1 \over 4} [(a_+ - a_-) x_0
+ (a_+ + a_-) x_3 ]^2 +  a_+ a_- x_1^2
\nonumber
\end{eqnarray}
One should have in mind $x_0 = 1$ for normalized density
operators. Then the last equation represents just Wootters
concurrence,
$C(\varrho)$, for density operators supported by $\cH_2$.
Returning to the amplitudes (\ref{e4.2}), $a$ and $b$,
results in a more convenient form
\begin{eqnarray}
F_{\Theta}(\varrho) &=& A
\sqrt{(x_0 - ab x_3)^2 - (1- a^2 b^2) x_2^2 },\\
C_{\Theta}(\varrho) &=& A
\sqrt{(x_3 - ab x_0)^2 + (1- a^2 b^2) x_1^2 },
\end{eqnarray}
\begin{equation}
A := {\sqrt{(1-a^2)(1-b^2)} \over 1- a^2 b^2}
\end{equation}
One easily determines to convex leaves for these roofs:
For $F_{\Theta}$ we fix $x_0=1$, $x_2$, and $x_3$ and let
$x_1$ vary. We obtain a straight line in $x$-space which
intersects the Bloch sphere of $\cH_2$ exactly twice,
corresponding to the two $x_3$-values with which
$x_1, x_2, x_3$ becomes a unit vector.
Along the line the $\Theta$-fidelity remains constant.\\
The same procedure, however with fixing $x_0, x_1, x_3$
and varying $x_2$, produces the convex foliation for
the $\Theta$-concurrence --- which, in our example, is the
Hill, Wootters one.

\acknowledgments
I like to thank P.~Alberti, B.~Crell, J.~Dittmann,
Ch.~Fuchs, R.~Jozsa, and W.~Wootters for valuable discussions.
Part of this work has been completed during the Newton Institute
workshop ''Computation, Complexity and the Physics of Information''
and the ESF-QIT programme meeting 1999 in Cambridge.
We acknowledge support of the European Science
Foundation QIT programme.


\begin{references}
\bibitem{T}
The implementation of time reversal in the algebra of quantum
fields is combined, by convention, with taking the Hermitian
adjoint, $A \to A^{\top} := \Theta A^{\dagger} \Theta$,
resulting in a transposition.
\bibitem{fuchs}
Independently this has been shown by Ch.~A.~Fuchs
with other techniques.
\end{references}
\end{document}